\documentclass[conference]{IEEEtran}
\usepackage{cite}
\usepackage{amsmath,amssymb,amsfonts}
\usepackage{algorithmic}
\usepackage{graphicx}
\usepackage{textcomp}
\usepackage{xcolor}
\def\BibTeX{{\rm B\kern-.05em{\sc i\kern-.025em b}\kern-.08em
    T\kern-.1667em\lower.7ex\hbox{E}\kern-.125emX}}

\usepackage{cite}
\usepackage{orcidlink} 
\usepackage{amsmath,amssymb,amsfonts}
\usepackage{xcolor}
\usepackage{algorithmic}
\usepackage{algorithm}
\usepackage{array}
\usepackage[caption=false,font=normalsize,labelfont=sf,textfont=sf]{subfig}
\usepackage{textcomp}
\usepackage{stfloats}
\usepackage{url}
\usepackage{verbatim}
\usepackage{graphicx}
\usepackage{cite}
\usepackage{siunitx}
\usepackage{graphicx}
\usepackage{amsmath, amsthm, amssymb, bm}
\usepackage{hyperref}
\usepackage{academicons}
\usepackage{xcolor}
\newcommand{\orcid}[1]{\href{https://orcid.org/#1}{\textcolor[HTML]{A6CE39}{\aiOrcid}}}
\graphicspath{{./images/}}
\newcommand{\eqnref}[1]{(\ref{#1})}
\newcommand{\figref}[1]{Figure~\ref{#1}}

\newcommand{\secref}[1]{Section~\ref{#1}}
\begin{document}

\title{Sub-band Domain Multi-Hypothesis  Acoustic Echo Canceler Based Acoustic Scene Analysis}

\author{
	\IEEEauthorblockN{Benjamin J. Southwell \orcidlink{0000-0002-3189-9114}}
	\IEEEauthorblockA{
		\textit{Dolby Laboratories}\\
		Sydney, Australia  \\
		benjamin.southwell@dolby.com}
	\and 
	\IEEEauthorblockN{Yin-Lee Ho \orcidlink{0009-0002-4648-2204}}
	\IEEEauthorblockA{
		\textit{Dolby Laboratories}\\
		Sydney, Australia  \\
		yin-lee.ho@dolby.com}
	\and
	\IEEEauthorblockN{David Gunawan \orcidlink{0009-0004-8606-3842}}
\IEEEauthorblockA{ 
	\textit{Dolby Laboratories}\\
	Sydney, Australia  \\
	david.gunawan@dolby.com}
}

\maketitle

\begin{abstract}
	This paper introduces a novel approach for acoustic scene analysis by exploiting an ensemble of statistics extracted from a sub-band domain multi-hypothesis acoustic echo canceler (SDMH-AEC). 
	A well-designed SDMH-AEC employs multiple adaptive filtering strategies with potentially complementary behaviors during convergence, perturbations, and steady-state conditions. 
	By aggregating statistics across the sub-bands, we derive a feature vector that exhibits strong discriminative power for distinguishing different acoustic events and estimating acoustic parameters. The complementary nature of the SDMH-AEC filters provides a rich source of information that can be extracted at insignificant cost for acoustic scene analysis tasks. 
	We demonstrate the effectiveness of the proposed approach experimentally with real data containing double-talk, echo path change  and  events where the full-duplex device is physically moved. The extracted features enable acoustic scene analysis using existing echo cancellation algorithms and techniques.
\end{abstract}


\section{Introduction}
Acoustic echo cancellation (AEC) can be performed using an adaptive filter comprised of taps $\hat{\bm{h}}$ that produces a prediction, $\hat{y}$, of the actual echo, $y$, in the microphone feed, $d$, using a delay line of a reference signal, $\bm{x}$ which is played by a speaker into the acoustic space shared by the microphone. Using the adaptive finite impulse response filter, the predicted signal at time $n$ is
\begin{equation}
	\hat{y}\left[n\right] = \bm{x}^T\left[n\right] \hat{ \bm{h}}\left[n\right] 
	\label{eq:predict}
\end{equation}
The residual signal, $e\left[n\right] = d\left[n\right]  - \hat{y}\left[n\right]$ is used to update the adaptive filter.
The normalized least-mean-square (NLMS) update to the adaptive coefficients is given by  \cite{haykin2002adaptive}
\begin{equation}
	\hat{ \bm{h}}\left[n+1\right] = \hat{ \bm{h}}\left[n\right] + \mu \frac{e^*\left[n\right]  \bm{x} \left[n\right] }{\bm{x}^T \left[n\right]\bm{x} \left[n\right]}
	\label{eq:nlms_algorithm}
\end{equation}
where the $*$ superscript denotes complex conjugation, a $T$ superscript denotes a transpose, and $\mu$ is the step size chosen by the designer.
The microphone signal can be modeled as $d\left[n\right]  =  \bm{x}^T\left[n\right]  {\bm{h}}\left[n\right]  +v\left[n\right]$ where ${\bm{h}}$ is the actual impulse response between the speaker and microphone, $v$ is noise, including user speech or double-talk, that perturbs the adaptive filter update \eqref{eq:nlms_algorithm}.

The length of the  impulse response, $\bm{h}$ contributing significant echo $y\left[n\right]  = \bm{x}^T\left[n\right]  {\bm{h}}\left[n\right] $  to the microphone feed can be on the order of 100s of milliseconds requiring  $\hat{ \bm{h}}$  to be potentially that long to model a sufficient amount of $\bm{h}$.
For example, with a sampling rate of 48kHz and a desired filter length of 200ms, we require $ \hat{ \bm{h}}$ to have 9600 elements which incurs significant computational cost to implement the convolution in \eqref{eq:predict}.
Moreover, consecutive reference delay line states, e.g. $\bm{x}\left[n\right]$ and $\bm{x}\left[n+1\right]$, exhibit high correlation for typical audio signals such as speech and music, leading to poor convergence rates \cite{haykin2002adaptive, albu2004combined}.
Augmentations to the learning algorithm, e.g. the affine projection filter \cite{ozeki1984adaptive}, can mitigate the poor convergence rate at the expense of additional compute. 
Adaptive filters operating on critically sampled sub-bands provide computational savings and can also be seen as a pre-whitening process \cite{gilloire1992adaptive, haykin2002adaptive, lee2009subband} improving the convergence rate.

In a practical system, we expect non-stationarity in both $v$ and $\bm{h}$. 
To mitigate this, AECs can be implemented with two echo path models \cite{albu2004combined, ochiai1977echo, lindstrom2007improvement}.
One filter, often called a background filter, is allowed to continually adapt, while heuristics copy its coefficients into a foreground filter that produces the output residual signal.
This can be generalized to $N$ echo path models and if applied in the sub-band domain yields the sub-band domain multi-hypothesis AEC (SDMH-AEC), e.g. \cite{nosrati2023learnable}.

The designer of an MH-AEC typically chooses filter adaption strategies that are complimentary. 
For example, for $N=2$, one filter may be aggressive to provide faster convergence and the other more conservative in order to avoid divergence during perturbations.
Control heuristics are built with prior knowledge of the filters characteristics and are responsible for resetting and copying coefficients between the $N$ filters.
The filter adaption strategies may differ in many ways, including the core adaption algorithm where NLMS, proportionate NLMS (PNLMS) \cite{Duttweiler2000}, the improved PNLMS \cite{benesty2002improved} and the Gauss-Seidel method  \cite{albu2002gauss, albu2004combined} are a few notable examples.
Additionally, numerous variable step size (VSS) algorithms, e.g. \cite{Kwong1992, mader2000step, Benesty2006, Ni2010, Grant2008, valin2007adjusting, Paleologu2008, Costa2008, Yim2015, Xia2016}, have been proposed which are designed to control the adaption by scaling the value of $\mu$ by some factor $F$.
Computing $F$ can require additional signal processing and estimation, sometimes with significant computational cost, that is not already done by the AEC.
Additionally, explicit acoustic state estimation can be performed entirely for the purpose of calculating $F$, e.g.  \cite{gansler1996double, benesty2000new,  schuldt2012delay}.
VSS schemes are often designed so that $F \to 1$ when the microphone contains only echo, and $F \to 0$ as more noise is present, addressing the trade-off between convergence rate and final misalignment associated with choosing a static $\mu$.
This improves robustness as the filter adaption is dampened during low signal to noise ratio conditions which may be caused by some perturbations such as double-talk or other sources of noise.

Acoustic scene analysis, especially noise estimation \cite{ris2001assessing, cohen2003noise, rangachari2006noise} and sound detection  \cite{chang2006voice, zhuang2010real, mesaros2021sound}  are often applied to contextually control echo management systems\cite{gansler1996double, benesty2000new,  schuldt2012delay, dickins2024acoustic}, the playback of interactive full duplex systems \cite{southwell2023ducking} and speech enhancement algorithms \cite{loizou2007speech} for example.
Techniques based on convolutional neural networks \cite{piczak2015environmental} or transformers \cite{gong2022ssast} which are computationally expensive may be employed for analysis. 
Other methods may even augment the playback signal in order to support acoustic scene analysis, e.g.  \cite{hines2024insertion}.
In this paper, we introduce a novel approach for acoustic scene analysis by exploiting an ensemble of statistics extracted from a SDMH-AEC that exhibits sensitivity to the acoustic scene without introducing any significant additional computational cost.

\section{A  multi-hypothesis canceler}
\begin{figure}
	\begin{center}
		{\includegraphics[width=\columnwidth]{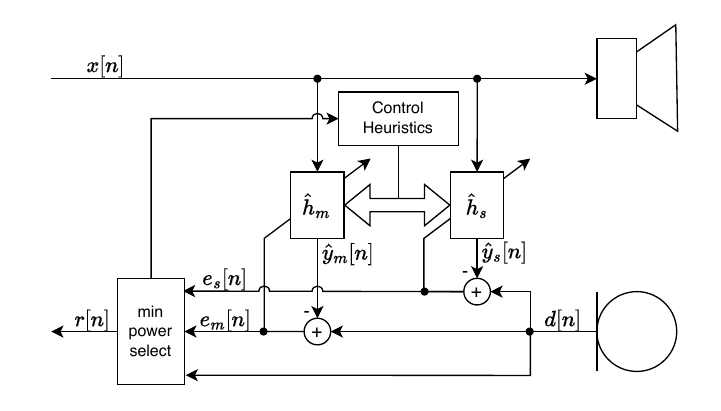}}
		\caption{A Multi-hypothesis AEC. In the sub-band scheme, this represents the processing that is done on a single sub-band where the analysis and synthesis transforms are not shown.}
		\label{fig:aec_multihypothesis}
	\end{center}
\end{figure}

In this section, we describe a SDMH-AEC, depicted in \figref{fig:aec_multihypothesis}, that copies coefficients between two active adaptive filters referred to as the main and  shadow filter which are designed to have complimentary adaption strategies.
The main filter utilizes a PNLMS adaption strategy with an update rate of $\mu_{main}=0.5$. 
The shadow filter utilizes a VSS-NLMS adaption strategy with a VSS of
\begin{equation}
	\mu_{shadow} =   \min\left( \frac{\left| \hat{y}_s\left[n\right] \right|^2}{\left| e_s\left[n\right] \right|^2}, 0.5 \right)
	\label{eq:vss_dt_scheme}
\end{equation}\\
Upon inspection of \eqnref{eq:vss_dt_scheme}, we can see that when the  predicted echo signal is high relative to the residual signal, the shadow filter will adapt at its quickest.
The shadow filter is conservative relative to the main filter.

The control heuristics block depicted in \figref{fig:aec_multihypothesis} will copy the main filter's coefficients into the shadow if  $\left| e_s\left[n\right] \right|^2$ is $\ge$10dB  more than $\left| e_m\left[n\right] \right|^2$ and vice versa to copy coefficients from the shadow into the main. 
An asymmetric holdover time is applied where the shadow must outperform the main for only 2 consecutive frames before copying will occur while the main must outperform the shadow for 5 before copying will occur.
As both filters are active, we expect them both to produce good residuals that are candidate for the overall output residual signal, $r\left[n\right]$, which is chosen from either the main residual, $e_m\left[n\right]$, the shadow residual, $e_s\left[n\right]$, or the microphone signal, $d\left[n\right]$, based on whichever has the lowest power.

\section{Extracting Statistics}
In the SDMH-AEC, we are processing $N_B$ sub-bands per microphone.
In this paper, we only consider the single microphone use case and we also do not explore the potential to extract statistics on a per-reference basis for multi-channel playback systems.
We extract a vector, $\bm{s} \in \mathbb{R}^5$, containing statistics aggregated across an ensemble of $N_S \le N_B$ sub-bands. 
The first three elements of $\bm{s}$ contain the probabilities of the main residual, shadow residual or microphone signal having the lowest power which is determined by the minimum power select block depicted in \figref{fig:aec_multihypothesis}.
That is,
\begin{equation}
	P_{m} = \frac{C_m}{N_S} 
	\text{,}\quad 
	P_{s} = \frac{C_s}{N_S} 
	\quad\text{and}\quad 
	P_{d} = \frac{C_d}{N_S}
\end{equation}
where $P_{m} + P_{s}  +  P_{d} = 1$.
$C_m$, $C_s$ and $C_d$ are the counts of the main residual, shadow residual and microphone signal having the lowest power respectively over the $N_S$ sub-bands.
The final two elements of $\bm{s}$ contain the probability, $U_{m}$, that the main filter has been updated with shadow filter coefficients and the probability, $U_{s},$ that the shadow filter has been updated with the main filter coefficients.
These are also computed by counting the number of occurrences over the $N_S$ sub-bands and normalizing it by $N_S$.
For the results shown here, we use an exponential moving average to smooth the statistics vector.
\begin{equation}
	\bm{\hat{s}}\left[n\right] = \alpha_s\bm{\hat{s}}\left[n-1\right] + (1-\alpha_s)\bm{s}\left[n\right]
\end{equation}
where  $\alpha_s = exp(-t_f / t_c)$; $t_f $ is the period of one frame of audio and $t_c$ is the smoothing time constant applied. For all statistics collected in this paper, we set $t_c = 200$ms. 
We run the canceler on data sampled at 48kHz and use a frame size of 512 samples, i.e. $t_f = 10.67$ms. 
Furthermore, we use a modified discrete Fourier transform \cite{karp1999modified} with modulated basis functions to produce 512 distinct sub-bands per frame.
However, we only use the lower $N_S = 100$ sub-bands which band-limits the statistics we aggregate from 0 Hz to 4687.5 Hz.

\section{Experimental Results}
\label{sec:statistics}
The results collected in this paper were produced using a full duplex device playing spectrally rich music. 
We truncate the beginning of all  the AEC outputs to ignore the initial convergence period.

\subsection{Steady-state operation}
\begin{figure}
	\begin{center}
		\includegraphics[width=0.825\columnwidth]{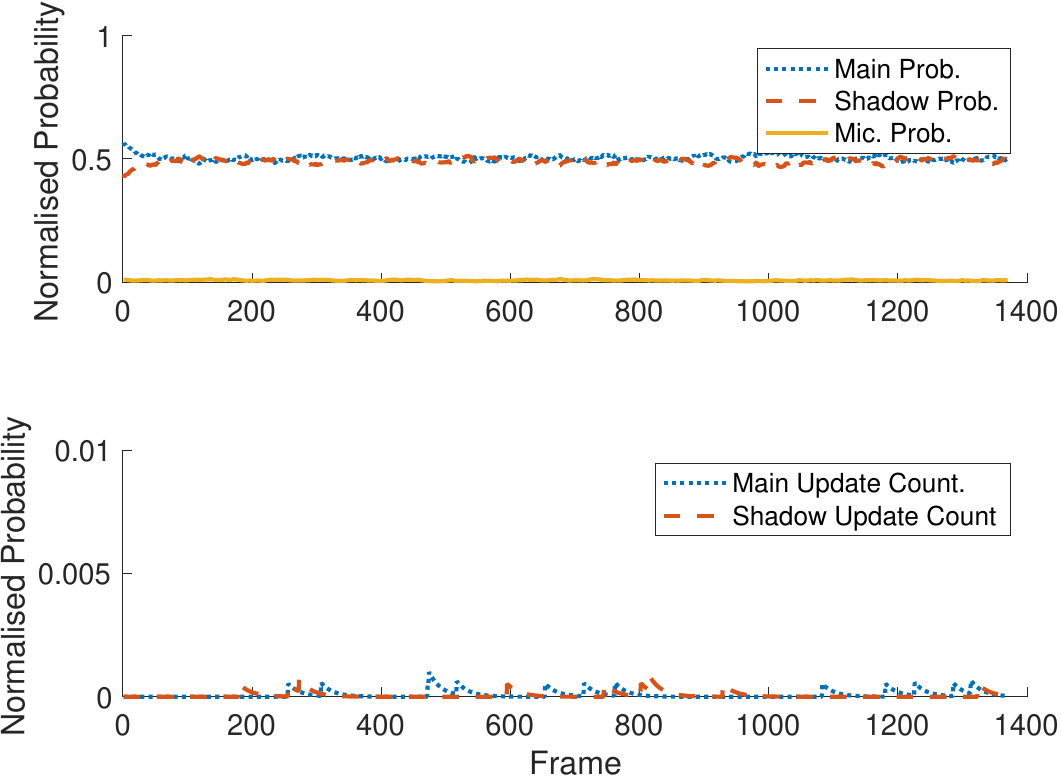}
		\caption{Elements of the statistics vector, $\bm{\hat{s}}\left[n\right] $, during steady-state operation with no perturbation. }
		\label{fig:silence}
	\end{center}
\end{figure}

In \figref{fig:silence},  $\bm{\hat{s}}\left[n\right] $ during steady-state operation canceling spectrally rich music with no perturbation is shown. 
We can see that the main and shadow filter both have a probability of approximately 0.5.
This is characteristic of the filters themselves and the operating conditions. 
The microphone probability is close to zero and we observe an insignificant amount of main and shadow update events.

\subsection{Double-Talk}
\begin{figure}
	\begin{center}
		\includegraphics[width=0.825\columnwidth]{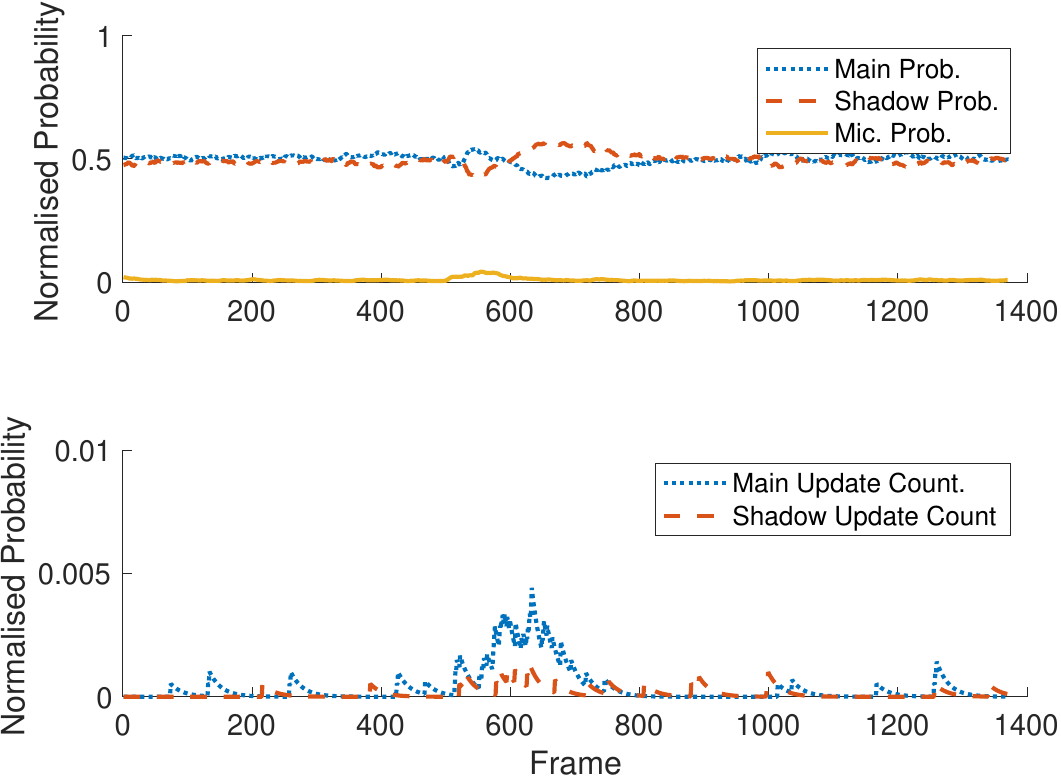}
		\caption{Elements of the statistics vector, $\bm{\hat{s}}\left[n\right] $, during steady-state operation with a double-talk event occurring at approximately frame 500.}
		\label{fig:double_talk}
	\end{center}
\end{figure}

In \figref{fig:double_talk}, $\bm{\hat{s}}\left[n\right] $ is shown with a double-talk event occurring at approximately frame 500.
We can see that before this, the statistics are similar to those shown in  \figref{fig:silence} as the AEC is operating in steady-state conditions.
As the double-talk event begins, $P_m$ momentarily increases as the main filter misadapts on the residual perturbed by the double-talk.
At this time, the shadow filter is not adapting due to the VSS \eqnref{eq:vss_dt_scheme} and the main filter produces a lower residual power as it has misadapted and is canceling some of the speech signal in $d\left[n\right]$.
Not long after this, $P_s$ increases and surpasses $P_m$ until approximately frame 800.
This is due to the main filter misadapting to the speech signal during the initial part of the perturbation.
After the misadaption, the shadow filter performs better than the main filter.
It does so by more than 10dB across  a number of sub-bands as we can see a significant level of $U_m$ beginning at the same frame.
Afterwards $U_m$ starts to fall-off, we can see that the AEC returns to steady state operation by frame 1000 approximately.

\subsection{Extrinsic echo path change}
\begin{figure}
	\begin{center}
		\includegraphics[width=0.825\columnwidth]{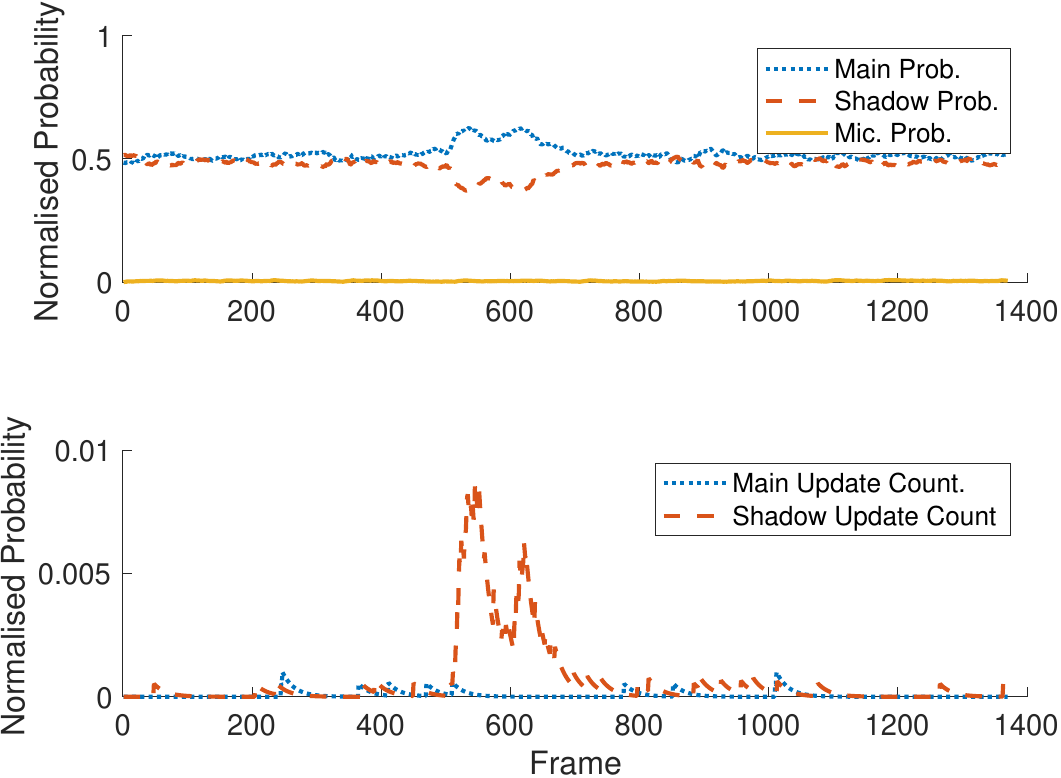}
		\caption{Elements of the statistics vector, $\bm{\hat{s}}\left[n\right] $, during steady-state operation with an echo path change event occurring at approximately frame 500.  }
		\label{fig:echo_path_change}
	\end{center}
\end{figure}

For the results shown in this paper, an extrinsic echo path change  perturbation was introduced by moving either a backpack or a laptop close to the full-duplex device for approximately 1 second (96 frames) and then moving it away.
In \figref{fig:echo_path_change}, $\bm{\hat{s}}\left[n\right] $ is shown with an extrinsic echo path change event occurring at approximately frame 500.
During the immediate onset of the event, we see that the main filter outperforms the shadow significantly as $P_m$ peaks.
Then we see a second peak not long after this.
These two peaks correspond to the impeding object being moved close to and then away from the device.
Both of these correspond to actual echo path change events as the echo path changes from the initial path to the one with the impeding device nearby and then back to the initial path.
We can see that there are corresponding dips in $P_s$ that occur concurrently with the $P_m$ peaks with no significant increase in $P_d$.
This indicates that the main filter is tracking the echo path changes well.
Immediately after a change in the echo path, the shadow filter estimates are incorrect and there is an increase in $\left| e_s\left[n\right] \right|^2$ and, similarly to the double-talk case, this causes the VSS scheme \eqnref{eq:vss_dt_scheme} to dampen the adaption of the shadow filter.
We observe a significant level of $U_s$ with two peaks corresponding to the two echo path change steps as the main filter is tracking the changes while the shadow is not.

\subsection{Device repositioning}
For the results shown in this paper, a device repositioning perturbation was introduced by momentarily picking up the full-duplex device for approximately 1 second (96 frames) and then putting it back in its original position.
Due to the coupling of the microphone to the chassis of the full-duplex device,  in addition to an echo path change which occurs due to the device moving, there is additional noise, $v\left[n\right]$, introduced into $d\left[n\right]$ when the user's hands initially grasp the device and later when the device is placed back onto its original position.
As a result, we expect this event to share features with both echo path change and double-talk events.

\begin{figure}
	\begin{center}
		\includegraphics[width=0.825\columnwidth]{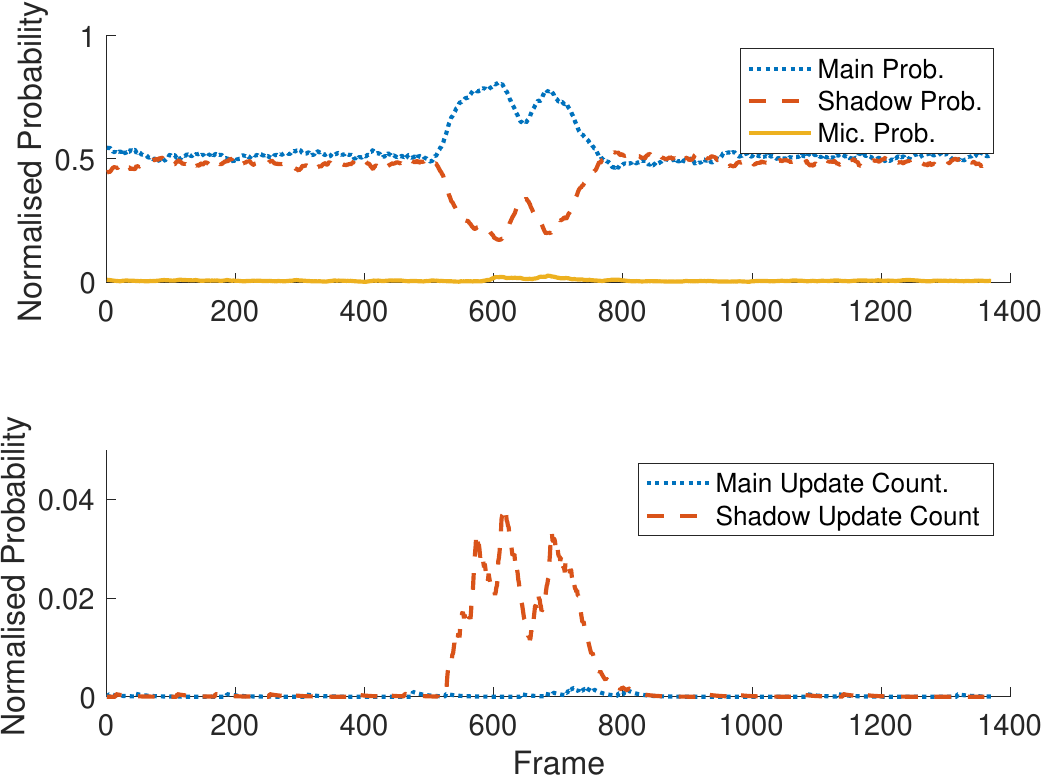}
		\caption{Elements of the statistics vector, $\bm{\hat{s}}\left[n\right] $, during steady-state operation with an event where the device is repositioned at approximately frame 500. }
		\label{fig:pickup_put_down}
	\end{center}
\end{figure}

In \figref{fig:pickup_put_down}, $\bm{\hat{s}}\left[n\right] $ is shown with a device repositioning event occurring at approximately frame 500.
We can observe a similar signal across $\bm{\hat{s}}\left[n\right]$ as that in \figref{fig:echo_path_change} where the echo path change event resulted in a double-peak in both $P_m$ and $U_s$ with a corresponding trough in $P_s$. 
Moreover, we also observe a peak in $U_m$ towards the end of the transient period where momentarily  $P_s > P_m$.
This is due to the noise in the microphone signal caused by placing the device back onto the surface upon which it is resting.
Noting the scale of the right-hand y axis in  \figref{fig:pickup_put_down} compared to the previous figures, we observe $U_m$ is on the same order of magnitude as that occurring in the echo path change event in \figref{fig:double_talk}.
There is significantly higher $U_s$ in \figref{fig:pickup_put_down} compared to \figref{fig:echo_path_change} as the step change in the echo path is larger due to the device actually moving rather than an object partially occluding a subset of the initial echo path.

\subsection{Acoustic scene analysis}
\begin{figure}
	\begin{center}
		\includegraphics[width=0.85\columnwidth]{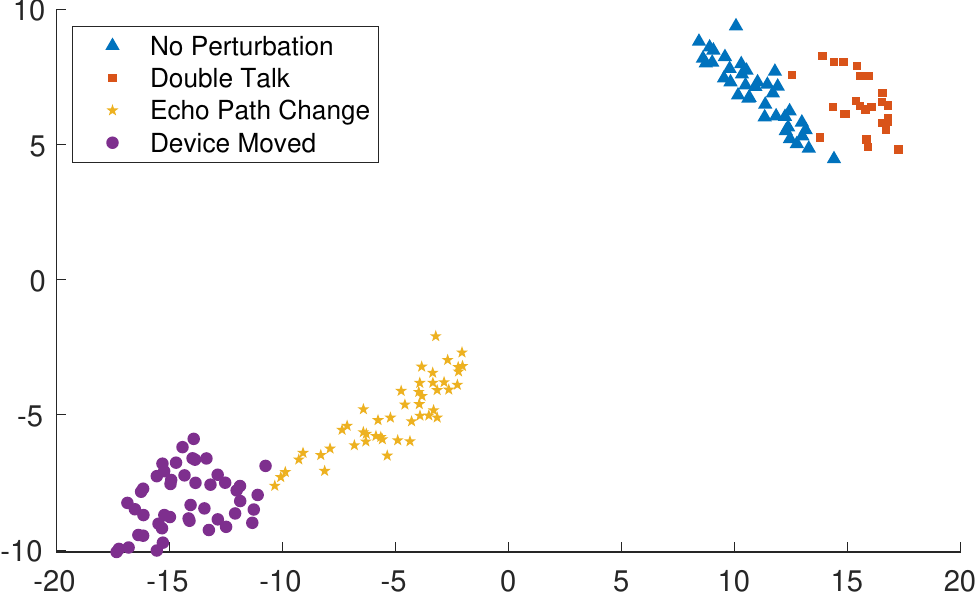}
		\caption{The two component t-SNE projection of $\bm{s}^\prime$: the features extracted from each statistics vector $\bm{s}\left[n\right]$ across all 4 classes. }
		\label{fig:tsne}
	\end{center}
\end{figure}

Thus far,  the statistics vector for 4 different acoustic events were analyzed and it was observed that each had a unique feature vector.
Here, we will perform an analysis of the statistics collected from 184 acoustic events across the same 4 classes.
A simple feature extraction method was devised where the mean, variance and dynamic range of each element of the statistic is aggregated over time into a vector $\bm{s}^\prime \in \mathbb{R}^{15}$.
\figref{fig:tsne} shows the t-SNE \cite{van2008visualizing}  projections of $\bm{s}^\prime$ down to 2 components.
We observe good separability between all classes while those which are similar in nature, e.g. echo path change and device repositioning, are closer together.
This suggests that $\bm{\hat{s}}\left[n\right]$ is a strong predictor for the classes we have analyzed in this paper. 

\section{Discussion}
The ensemble statistics that can be extracted from the metadata within an SDMH-AECs come at insignificant cost as they are simply probed from the already existing architecture required to implement the AEC.
Nevertheless, they have strong discriminatory power as they are derived from a multitude of diverse adaptive filtering algorithms and VSS schemes.
Some AECs have explicit state detectors, e.g. double-talk \cite{gansler1996double, benesty2000new,  schuldt2012delay}, constructed for the purposes of implementing a VSS for an adaptive filter.
It is possible to use the statistics gathered from a subset of the filters in the SDMH-AEC and use those to drive, e.g. with VSS, another subset of filters, either with or without, explicit state detection in the space connecting them.
Moreover, the ensemble statistics can provide additional state estimation capability.
For example \cite{southwell2023subband} showed that shifts in the noise floor of the system can be detected and tracked via the characteristic relative main and shadow probabilities.

The statics presented in \secref{sec:statistics} were collected by averaging over a range of the sub-bands to produce a single ensemble average across this one band of frequencies.
The use of banding strategies, e.g. mel-spaced banding, transforms, e.g. computing cepstrum coefficients, and preserving temporal information is well known to affect the ability of audio classifiers \cite{mckinney2003features, malik2021automatic}.
Future work will investigate the use of more complex feature vectors  as inputs to both classifiers and regressors for the purposes of detecting and estimating acoustic state.

\section{Conclusion}
This work introduced a novel approach to acoustic scene analysis by exploiting an ensemble of statistics extracted from a sub-band domain multi-hypothesis acoustic echo canceler (SDMH-AEC). 
By probing metadata such as filter coefficient copying rates and the relative performance of the complementary filtering strategies, we derived  statistics vector that demonstrated discriminative power across multiple acoustic states. 
An advantage of this approach is that there is insignificant cost associated with obtaining these statistics.
Future work will investigate the use of richer features and downstream classifiers and regressors for acoustic scene analysis.

\ifCLASSOPTIONpeerreview
\pagebreak
\fi

\bibliographystyle{IEEEtran}
\bibliography{IEEEabrv,references}

\begin{thebibliography}{10}
\providecommand{\url}[1]{#1}
\csname url@samestyle\endcsname
\providecommand{\newblock}{\relax}
\providecommand{\bibinfo}[2]{#2}
\providecommand{\BIBentrySTDinterwordspacing}{\spaceskip=0pt\relax}
\providecommand{\BIBentryALTinterwordstretchfactor}{4}
\providecommand{\BIBentryALTinterwordspacing}{\spaceskip=\fontdimen2\font plus
\BIBentryALTinterwordstretchfactor\fontdimen3\font minus
  \fontdimen4\font\relax}
\providecommand{\BIBforeignlanguage}[2]{{%
\expandafter\ifx\csname l@#1\endcsname\relax
\typeout{** WARNING: IEEEtran.bst: No hyphenation pattern has been}%
\typeout{** loaded for the language `#1'. Using the pattern for}%
\typeout{** the default language instead.}%
\else
\language=\csname l@#1\endcsname
\fi
#2}}
\providecommand{\BIBdecl}{\relax}
\BIBdecl

\bibitem{haykin2002adaptive}
S.~Haykin, \emph{Adaptive filter theory}, 5th~ed.\hskip 1em plus 0.5em minus
  0.4em\relax Upper Saddle River, NJ: Prentice Hall, 2014.

\bibitem{albu2004combined}
F.~Albu and H.~K. Kwan, ``Combined echo and noise cancellation based on
  {Gauss-Seidel} pseudo affine projection algorithm,'' in \emph{2004 IEEE
  International Symposium on Circuits and Systems}, vol.~3.\hskip 1em plus
  0.5em minus 0.4em\relax IEEE, 2004, pp. III--505--III--508.

\bibitem{ozeki1984adaptive}
K.~Ozeki and T.~Umeda, ``An adaptive filtering algorithm using an orthogonal
  projection to an affine subspace and its properties,'' \emph{Electronics and
  Communications in Japan (Part I: Communications)}, vol.~67, no.~5, pp.
  19--27, 1984.

\bibitem{gilloire1992adaptive}
A.~Gilloire and M.~Vetterli, ``Adaptive filtering in sub-bands with critical
  sampling: analysis, experiments, and application to acoustic echo
  cancellation,'' \emph{IEEE transactions on signal processing}, vol.~40,
  no.~8, pp. 1862--1875, 1992.

\bibitem{lee2009subband}
K.-A. Lee, W.-S. Gan, and S.~M. Kuo, \emph{Subband adaptive filtering: theory
  and implementation}.\hskip 1em plus 0.5em minus 0.4em\relax Chichester, U.K:
  John Wiley \& Sons, 2009.

\bibitem{ochiai1977echo}
K.~Ochiai, T.~Araseki, and T.~Ogihara, ``Echo canceler with two echo path
  models,'' \emph{IEEE Transactions on Communications}, vol.~25, no.~6, pp.
  589--595, 1977.

\bibitem{lindstrom2007improvement}
F.~Lindstrom, C.~Schuldt, and I.~Claesson, ``An improvement of the two-path
  algorithm transfer logic for acoustic echo cancellation,'' \emph{IEEE
  transactions on audio, speech, and language processing}, vol.~15, no.~4, pp.
  1320--1326, 2007.

\bibitem{nosrati2023learnable}
\BIBentryALTinterwordspacing
H.~Nosrati and B.~J. Southwell, ``Learnable heuristics to optimize a
  multi-hypothesis filtering system,'' WO Patent WO2\,023\,086\,244A1, May 19,
  2023. [Online]. Available:
  \url{https://patents.google.com/patent/WO2023086244A1}
\BIBentrySTDinterwordspacing

\bibitem{Duttweiler2000}
D.~L. Duttweiler, ``{Proportionate normalized least-mean-squares adaptation in
  echo cancelers},'' \emph{IEEE Transactions on Speech and Audio Processing},
  vol.~8, no.~5, pp. 508--517, 2000.

\bibitem{benesty2002improved}
J.~Benesty and S.~L. Gay, ``An improved {PNLM}s algorithm,'' in \emph{2002 IEEE
  international conference on acoustics, speech, and signal processing},
  vol.~2.\hskip 1em plus 0.5em minus 0.4em\relax IEEE, 2002, pp. II--1881.

\bibitem{albu2002gauss}
F.~Albu, J.~Kadlec, N.~Coleman, and A.~Fagan, ``The {Gauss-Seidel} fast affine
  projection algorithm,'' in \emph{IEEE Workshop on Signal Processing
  Systems}.\hskip 1em plus 0.5em minus 0.4em\relax IEEE, 2002, pp. 109--114.

\bibitem{Kwong1992}
R.~H. Kwong and E.~W. Johnston, ``{A Variable Step Size LMS Algorithm},''
  \emph{IEEE Transactions on Signal Processing}, vol.~40, no.~7, pp.
  1633--1642, 1992.

\bibitem{mader2000step}
A.~Mader, H.~Puder, and G.~U. Schmidt, ``Step-size control for acoustic echo
  cancellation filters - an overview,'' \emph{Signal Processing}, vol.~80,
  no.~9, pp. 1697--1719, 2000.

\bibitem{Benesty2006}
J.~Benesty, H.~Rey, L.~R. Vega, and S.~Tressens, ``{A nonparametric VSS NLMS
  algorithm},'' \emph{IEEE Signal Processing Letters}, vol.~13, no.~10, pp.
  581--584, 2006.

\bibitem{Ni2010}
J.~Ni and F.~Li, ``{A variable step-size matrix normalized subband adaptive
  filter},'' \emph{IEEE Transactions on Audio, Speech and Language Processing},
  vol.~18, no.~6, pp. 1290--1299, 2010.

\bibitem{Grant2008}
S.~L. Grant, ``Novel variable step size {NLMS} algorithms for echo
  cancellation,'' pp. 241--244, 2008.

\bibitem{valin2007adjusting}
J.-M. Valin, ``On adjusting the learning rate in frequency domain echo
  cancellation with double-talk,'' \emph{IEEE Transactions on Audio, Speech,
  and Language Processing}, vol.~15, no.~3, pp. 1030--1034, 2007.

\bibitem{Paleologu2008}
C.~Paleologu, S.~Ciochina, and J.~Benesty, ``{Variable step-size NLMS algorithm
  for under-modeling acoustic echo cancellation},'' \emph{IEEE Signal
  Processing Letters}, vol.~15, pp. 5--8, 2008.

\bibitem{Costa2008}
M.~H. Costa and J.~C.~M. Bermudez, ``{A noise resilient variable step-size LMS
  algorithm},'' \emph{Signal Processing}, vol.~88, no.~3, pp. 733--748, 2008.

\bibitem{Yim2015}
S.~H. Yim, H.~S. Lee, and W.~J. Song, ``A proportionate diffusion {LMS}
  algorithm for sparse distributed estimation,'' \emph{IEEE Transactions on
  Circuits and Systems II: Express Briefs}, vol.~62, no.~10, pp. 992--996,
  2015.

\bibitem{Xia2016}
W.~Xia, L.~Zhu, J.~Zhu, J.~Hu, and H.~Li, ``{A shrinkage variable step size for
  normalized subband adaptive filters},'' \emph{Signal Processing}, vol. 129,
  no.~12, pp. 56--61, 2016.

\bibitem{gansler1996double}
T.~Gansler, M.~Hansson, C.-J. Ivarsson, and G.~Salomonsson, ``A double-talk
  detector based on coherence,'' \emph{IEEE Transactions on Communications},
  vol.~44, no.~11, pp. 1421--1427, 1996.

\bibitem{benesty2000new}
J.~Benesty, D.~R. Morgan, and J.~H. Cho, ``A new class of doubletalk detectors
  based on cross-correlation,'' \emph{IEEE Transactions on Speech and Audio
  Processing}, vol.~8, no.~2, pp. 168--172, 2000.

\bibitem{schuldt2012delay}
C.~Schuldt, F.~Lindstrom, and I.~Claesson, ``A delay-based double-talk
  detector,'' \emph{IEEE transactions on audio, speech, and language
  processing}, vol.~20, no.~6, pp. 1725--1733, 2012.

\bibitem{ris2001assessing}
C.~Ris and S.~Dupont, ``Assessing local noise level estimation methods:
  Application to noise robust asr,'' \emph{Speech communication}, vol.~34, no.
  1-2, pp. 141--158, 2001.

\bibitem{cohen2003noise}
I.~Cohen, ``Noise spectrum estimation in adverse environments: Improved minima
  controlled recursive averaging,'' \emph{IEEE Transactions on speech and audio
  processing}, vol.~11, no.~5, pp. 466--475, 2003.

\bibitem{rangachari2006noise}
S.~Rangachari and P.~C. Loizou, ``A noise-estimation algorithm for highly
  non-stationary environments,'' \emph{Speech communication}, vol.~48, no.~2,
  pp. 220--231, 2006.

\bibitem{chang2006voice}
J.-H. Chang, N.~S. Kim, and S.~K. Mitra, ``Voice activity detection based on
  multiple statistical models,'' \emph{IEEE Transactions on Signal Processing},
  vol.~54, no.~6, pp. 1965--1976, 2006.

\bibitem{zhuang2010real}
X.~Zhuang, X.~Zhou, M.~A. Hasegawa-Johnson, and T.~S. Huang, ``Real-world
  acoustic event detection,'' \emph{Pattern recognition letters}, vol.~31,
  no.~12, pp. 1543--1551, 2010.

\bibitem{mesaros2021sound}
A.~Mesaros, T.~Heittola, T.~Virtanen, and M.~D. Plumbley, ``Sound event
  detection: A tutorial,'' \emph{IEEE Signal Processing Magazine}, vol.~38,
  no.~5, pp. 67--83, 2021.

\bibitem{dickins2024acoustic}
\BIBentryALTinterwordspacing
G.~N. Dickins, C.~G. Hines, D.~Gunawan, R.~J. Cartwright, A.~J. Seefeldt,
  D.~Arteaga, M.~R. Thomas, and J.~B. Lando, ``Acoustic echo cancellation
  control for distributed audio devices,'' US Patent 17,628,732, Jun., 2024.
  [Online]. Available: \url{https://patents.google.com/patent/US12003673B2/en}
\BIBentrySTDinterwordspacing

\bibitem{southwell2023ducking}
\BIBentryALTinterwordspacing
B.~J. Southwell, D.~Gunawan, and A.~J. Seefeldt, ``Distributed audio device
  ducking,'' WO Patent PCT/US2022/048\,956, May 19, 2023. [Online]. Available:
  \url{https://patents.google.com/patent/WO2023086273A1/en}
\BIBentrySTDinterwordspacing

\bibitem{loizou2007speech}
P.~C. Loizou, \emph{Speech enhancement: theory and practice}.\hskip 1em plus
  0.5em minus 0.4em\relax Boca Raton: CRC press, 2007.

\bibitem{piczak2015environmental}
K.~J. Piczak, ``Environmental sound classification with convolutional neural
  networks,'' in \emph{2015 IEEE 25th international workshop on machine
  learning for signal processing (MLSP)}.\hskip 1em plus 0.5em minus
  0.4em\relax IEEE, 2015, pp. 1--6.

\bibitem{gong2022ssast}
Y.~Gong, C.-I. Lai, Y.-A. Chung, and J.~Glass, ``{SSAST}: Self-supervised audio
  spectrogram transformer,'' in \emph{Proceedings of the AAAI Conference on
  Artificial Intelligence}, vol.~36, no.~10, 2022, pp. 10\,699--10\,709.

\bibitem{hines2024insertion}
\BIBentryALTinterwordspacing
C.~G. Hines and B.~J. Southwell, ``Insertion of forced gaps for pervasive
  listening,'' US Patent 18/254,962, Mar. 28, 2024. [Online]. Available:
  \url{https://patents.google.com/patent/US20240107252A1/en}
\BIBentrySTDinterwordspacing

\bibitem{karp1999modified}
T.~Karp and N.~J. Fliege, ``Modified {DFT} filter banks with perfect
  reconstruction,'' \emph{IEEE Transactions on Circuits and Systems II: Analog
  and Digital Signal Processing}, vol.~46, no.~11, pp. 1404--1414, 1999.

\bibitem{van2008visualizing}
L.~Van~der Maaten and G.~Hinton, ``Visualizing data using {t-SNE}.''
  \emph{Journal of machine learning research}, vol.~9, no.~11, pp. 2579--2605,
  2008.

\bibitem{southwell2023subband}
\BIBentryALTinterwordspacing
B.~J. Southwell, D.~Gunawan, and C.~G. Hines, ``Subband domain acoustic echo
  canceller based acoustic state estimator,'' US Patent 18/255,573, Dec. 28,
  2023. [Online]. Available:
  \url{https://patents.google.com/patent/US20230421952A1/en}
\BIBentrySTDinterwordspacing

\bibitem{mckinney2003features}
M.~F. McKinney and J.~Breebaart, ``Features for audio and music
  classification,'' in \emph{Proceedings of the International Society for Music
  Information Retrieval Conference (ISMIR)}, 2003, pp. 151--158.

\bibitem{malik2021automatic}
M.~Malik, M.~K. Malik, K.~Mehmood, and I.~Makhdoom, ``Automatic speech
  recognition: a survey,'' \emph{Multimedia Tools and Applications}, vol.~80,
  pp. 9411--9457, 2021.

\end{thebibliography}

\vspace{12pt}

\end{document}